\pdfoutput=1
\documentclass[10.75pt, a4paper]{article}
\usepackage{titlesec}
\titleformat{\section}
  {\bf\sffamily}
  {\thesection. }
  {5pt}
  {\MakeUppercase}
\renewcommand{\thesection}{\Roman{section}} 

\titleformat{\subsection}
  {\bf\sffamily}
  {\thesubsection. }
  {5pt}{}
  
\renewcommand{\thesubsection}{\Alph{subsection}}

\usepackage[affil-it]{authblk} 
\usepackage{etoolbox}
\usepackage{lmodern}

\usepackage{eurosym}

\usepackage[utf8]{inputenc}
\usepackage{geometry}
\usepackage[hidelinks]{hyperref}
\usepackage[citestyle=numeric,style=phys,biblabel=brackets,backend=bibtex,sorting=none,url=false,doi=false, natbib]{biblatex}
\bibliography{main}
\DefineBibliographyStrings{english}{andothers={\itshape et\addabbrvspace al\adddot}}

\usepackage{csquotes}
\usepackage[]{authblk} 
\usepackage{etoolbox}
\usepackage{lmodern}

\usepackage{amsmath}
\usepackage{enumerate}
\usepackage{enumitem}
\usepackage{graphicx}
\usepackage{siunitx}
\usepackage{float}
\usepackage[]{parskip} %skip=0.7em, indent=1.5em

\makeatletter
\patchcmd{\@maketitle}{\LARGE \@title}{\fontsize{16}{19.2}\selectfont\@title}{}{}
\makeatother

\usepackage[normalem]{ulem}

\usepackage{graphicx}% Include figure files
\usepackage{dcolumn}% Align table columns on decimal point
\usepackage{bm}% bold math
\usepackage{tikz}
\usetikzlibrary{math}
\usetikzlibrary{shapes.geometric, arrows}
\usetikzlibrary{positioning}
\usetikzlibrary{shapes,arrows}
\usetikzlibrary{intersections}
\usepackage{csvsimple}
\usepackage{changepage}
\usepackage[tbtags]{mathtools}
\usepackage{siunitx}
\usepackage{csquotes}

\usepackage{float}
\usepackage{subfigure}
\usepackage[
	nonumberlist, 				% keine Seitenzahlen anzeigen
	acronym,      				% ein Abk�rzungsverzeichnis erstellen
	nomain,						% no main glossary
	nopostdot,					% Kein punkt nach Beschreibung
]{glossaries}
\usepackage{glossary-superragged}
\usepackage[hidelinks]{hyperref}
\usepackage{color, colortbl}

\usepackage{wrapfig}

\usepackage{pgfplots}
\usepackage{tikz}
\usetikzlibrary{arrows}
\usepackage{amsmath}
\pgfplotsset{compat=newest}
\usepgfplotslibrary{fillbetween}
\usetikzlibrary{calc}
\def\centerarc[#1](#2)(#3:#4:#5)% Syntax: [draw options] (center) (initial angle:final angle:radius)
{ \draw[#1] ($(#2)+({#5*cos(#3)},{#5*sin(#3)})$) arc (#3:#4:#5); }

\usepackage{amssymb}
\let\vec\mathbf
\usepackage{nicefrac}

\usepackage{abstract}
\usepackage{multirow}
\usepackage{tabularx}
\usepackage{booktabs}
\usepackage{array}
\usepackage{longtable}
\newcolumntype{L}[1]{>{\raggedright\let\newline\\\arraybackslash\hspace{0pt}}m{#1}}
\newcolumntype{C}[1]{>{\centering\let\newline\\\arraybackslash\hspace{0pt}}m{#1}}
\newcolumntype{R}[1]{>{\raggedleft\let\newline\\\arraybackslash\hspace{0pt}}m{#1}}

\newacronym{3d}{3D}{three dimensional}
\newacronym{am}{AM}{additive manufacturing}
\newacronym{fdm}{FDM}{fused deposition modeling}
\newacronym{ism}{ISM}{in-space manufacturing}
\newacronym{iss}{ISS}{International Space Station}
\newacronym{fcb}{FCB}{Functional Cargo Block}
\newacronym{dem}{DEM}{discrete element method}
\newacronym{md}{MD}{molecular dynamics}
\newacronym{dc}{DC}{direct-current}
\newacronym[plural=PFCs,firstplural=parabolic flight campaigns (PFCs)]{pfc}{PFC}{Parabolic Flight Campaign}
\newacronym{fft}{FFT}{Fast Fourrier Transform}
\newacronym{cad}{CAD}{Computer Assisted Design}
\newacronym{ptfe}{PTFE}{polytetrafluoroethylene}
\newacronym{ps}{PS}{polystyrene}
\newacronym{nasa}{NASA}{National Aeronautics and Space Administration}
\newacronym{esamm}{ESAMM}{Extended Structure Additive Manufacturing Machine}
\newacronym{amf}{AMF}{Additive Manufacturing Facility}
\newacronym{us}{US}{United States}
\newacronym{usa}{USA}{United States of America}
\newacronym{bmgs}{BMGs}{Bulk Metallic Glasses}
\newacronym{esa}{ESA}{European Space Agency}
\newacronym{si}{SI}{International System of Units, abbreviated from French \textit{Syst\`{e}me International (d'unit\'{e}s)}}
\newacronym{dlr}{DLR}{German Aerospace Center}
\newacronym{liggghts}{LIGGGHTS}{\acrshort{lammps} Improved for General Granular and Granular Heat Transfer Simulations}
\newacronym{lammps}{LAMMPS}{Large-scale Atomic/Molecular Massively Parallel Simulator}
\newacronym{sjkr}{SJKR}{Simplified Johnson-Kendall-Roberts}
\newacronym{ded}{DED}{Directed Energy Deposition}
\newacronym{slm}{SLM}{Selective Laser Melting}
\newacronym{sls}{SLS}{Selective Laser Sintering}
\newacronym{eva}{EVA}{Extra-Vehicular Activity}
\newacronym{sem}{SEM}{Scanning Electron Microscopy}
\newacronym{RPM}{RPM}{Ramdom Positioning Machine}
\newacronym{rpm}{rpm}{revolutions per minute}
\newacronym{rise}{RISE}{Research Internships in Science and Engineering}
\newacronym{daad}{DAAD}{German Academic Exchange Service, abbreviated from German \textit{Deutscher Akademischer Austauschdienst}}
\newacronym{fsm}{FSM}{finite-state machine}
\newacronym{ir}{IR}{infrared}
\newacronym{pcbs}{PCBs}{Printed Circuit Boards}
\newacronym{pcb}{PCB}{Printed Circuit Board}
\newacronym{mcr}{MCR}{Modular Compact Rheometer}
\newacronym{sff}{SFF}{Solid Freeform Fabrication}
\newacronym{uv}{UV}{ultraviolet}
\newacronym{abs}{ABS}{acrylonitrile butadiene styrene}
\newacronym{hpde}{HPDE}{high density polyethylene}
\newacronym{pei}{PEI}{polyetherimide}
\newacronym{bff}{BFF}{BioFabrication Facility}
\newacronym{lens}{LENS}{Laser Engineered Net Shaping}
\newacronym{cnc}{CNC}{Computer Numerical Control}
\newacronym{ebf3}{EBF$^3$}{Electron Beam Free-Form Fabrication}
\newacronym{leo}{LEO}{Low Earth Orbit}
\newacronym{pc}{PC}{polycarbonate}
\newacronym{crissp}{CRISSP}{Customisable Recyclable International Space Station Packaging}
\newacronym{Athena}{Athena}{Advanced Telescope for High-ENergy Astrophysics}
\newacronym{lbm}{LBM}{Laser Beam Melting}
\newacronym{bam}{BAM}{Federal Institute for Materials Research and Testing, abbreviated from German \textit{Bundesanstalt f\"{u}r Materialforschung und-pr\"{u}fung}}
\newacronym{pbf}{PBF}{powder bed fusion}
\newacronym{eb}{EB}{Electron Beam}
\newacronym{2d}{2D}{two dimensional}
\newacronym{4d}{4D}{four dimensional}
\newacronym{ft4}{FT4}{Freeman Technology 4 Powder Rheometer}
\newacronym{dsc}{DSC}{Differential Scanning Calorimetry}
\newacronym{pmma}{PMMA}{polymethylmethacrylate}
\newacronym{1g}{$1g$}{gravity on-ground}
\newacronym{mug}{$\mu g$}{microgravity}
\newacronym{bcm}{BCM}{Box Counting Method}
\newacronym{mct}{MCT}{Mode Coupling Theory}
\newacronym{gmct}{gMCT}{granular Mode Coupling Theory}
\newacronym{itt}{ITT}{Integration Through Transients}
\newacronym{mfc}{MFC}{Mass Flow Controller}
\newacronym{ct}{CT}{computed tomography}
\newacronym{xct}{XCT}{X-ray computed tomography}
\newacronym{cv}{CV}{curriculum vitae}
\newacronym{pi}{PI}{principal investigator}
\newacronym{osp}{OSP}{orthogonal superimposed perturbation}
\newacronym{npi}{NPI}{Network Partnering Initiative}
\newacronym{ecsat}{ECSAT}{European Centre for Space Applications and Telecommunications}
\newacronym{eac}{EAC}{European Astronaut Centre}
\newacronym{estec}{ESTEC}{European Space Research and Technology Centre}
\newacronym{fps}{fps}{frames per second}
\newacronym{pdf}{pdf}{probability density function}
\newacronym{al}{Al}{aluminium}
\newacronym{ss}{\textit{SS}}{\textit{Smooth Surface}}
\newacronym{rs}{\textit{RS}}{\textit{Rough Surface}}
\newacronym{rcp}{rcp}{random close packing}
\newacronym{iop}{IoP UvA}{Institute of Physics of the University of Amsterdam}
\newacronym{mp}{MP}{Institute of Material Physics for Space}
\newacronym{elgra}{ELGRA}{European Low Gravity Research Association}
\newacronym{zarm}{ZARM}{Center of Applied Space Technology and Microgravity}
\newacronym{piv}{PIV}{particle image velocimetry}
\usepackage[framemethod=tikz]{mdframed}
\usepackage{lipsum}
\usepackage{dirtytalk}
\usepackage{tcolorbox}
\PassOptionsToPackage{table}{xcolor}
\usepackage{caption}
\newtcolorbox{mybox}[1]{colback=green!6!white,colframe=black!75!black,fonttitle=\bfseries,title=#1}
\newtcolorbox{mybox2}{colback=red!5!white,colframe=red!75!black}

\usepackage{pifont}

\def\be{\begin{equation}}
\def\ee{\end{equation}}

\usepackage{biblatex}
\usepackage{soul}
\setstcolor{red}

\usepackage{hyperref}
\hypersetup{
   colorlinks,
   linkcolor={blue!50!black},%{red!80!black},
   citecolor={blue!50!black},
   urlcolor={blue!80!black}
} 
\definecolor{mycolor}{rgb}{0.122, 0.435, 0.698}

\title{Glandular Trichome Rupture in Tomato Plants is an Ultra-Fast \& Sensitive Defense Mechanism Against Insects}

\author[1]{Jared Popowski\footnote{j.s.popowski@uva.nl, ORCID: 0009-0001-7507-5651}}
\author[1]{Lucas Warma}
\author[2]{Alicia Abarca Cifuentes}
\author[2]{Petra Bleeker}
\author[1]{Maziyar Jalaal\footnote{m.jalaal@uva.nl, ORCID: 0000-0002-5654-8505}}

\affil[1]{Van der Waals-Zeeman Institute, Institute of Physics, \protect\\
University of Amsterdam, Science Park 904, Amsterdam, 1098XH, The Netherlands}
\affil[2]{Department of Plant Physiology, Green Life Sciences Research Theme, Swammerdam Institute for Life Sciences,\protect\\
University of Amsterdam, 1098 XH Amsterdam, The Netherlands}

\begin{document}
\definecolor{brickred}{rgb}{0.8, 0.25, 0.33}
\definecolor{darkorange}{rgb}{1.0, 0.55, 0.0}
\definecolor{persiangreen}{rgb}{0.0, 0.65, 0.58}
\definecolor{persianindigo}{rgb}{0.2, 0.07, 0.48}
\definecolor{cadet}{rgb}{0.33, 0.41, 0.47}
\definecolor{turquoisegreen}{rgb}{0.63, 0.84, 0.71}
\definecolor{sandybrown}{rgb}{0.96, 0.64, 0.38}
\definecolor{blueblue}{rgb}{0.0, 0.2, 0.6}
\definecolor{ballblue}{rgb}{0.13, 0.67, 0.8}
\definecolor{greengreen}{rgb}{0.0, 0.5, 0.0}
\begingroup
\sffamily
\date{}
\maketitle
\endgroup

\begin{abstract}
Trichomes, specialized hair-like structures on the surfaces of many plants, play a crucial role in defense against herbivorous insects. 
We investigated the biomechanics of type VI glandular trichomes in cultivated tomato (\textit{Solanum lycopersicum}) and its wild relative (\textit{Solanum habrochaites}).
Using micropipette force sensors and high-speed imaging, we uncovered the rupture mechanics underlying gland bursting, highlighting the small forces and short time-scales involved in this process. Additionally, we observed larvae of the Western flower thrips (\textit{Frankliniella occidentalis}), a major pest in tomato cultivation, inadvertently triggering trichome rupture and accumulating glandular secretions on their bodies. These findings demonstrate how rapid gland bursting and the fluid dynamics of glandular secretions act as an efficient and swift plant defense mechanism against insect herbivory.

\textbf{keywords: trichomes $|$ biomechanics $|$ plant-insect interactions $|$ plant physics} 

\end{abstract}

%%%%%%%%%%%%%%%%%%%%%%%%%%%%%%% Introduction %%%%%%%%%%%%%%%%%%%%%%%%%%%%%%%%%%%%%%%%%%%%%%
\section*{Introduction}
Plants and the herbivores that consume them have been locked in an evolutionary arms race for hundreds of millions of years \cite{Labandeira1997InsectStrategies,Sharma2021InsectPlantRelationship}. The fossil record shows evidence of plant tissue damage from insect predation essentially as far back as vascular plants have existed \cite{Boucot2010FossilCompendium, Labandeira1993InsectRecord, Perez-delaFuente2016ADiogenesi}.
As a result, plants have developed increasingly more sophisticated defense systems against insects over time. 

\begin{figure}[!tbph]
    \centering
    \includegraphics[width=1\textwidth]{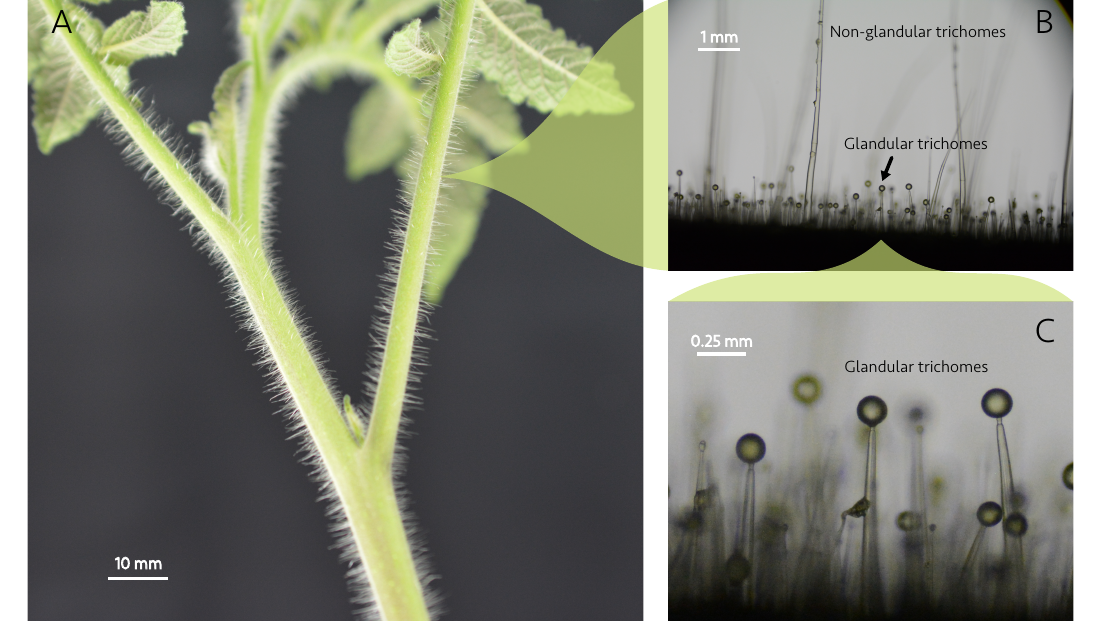}
    \caption{\textbf{The multiscale nature of trichomes.} \textbf{A.} The stem and leaves of a \textit{S. habrochaites} tomato plant, where the largest non-glandular trichomes are clearly visible as hair-like protrusions at distance from the plant surface. \textbf{B.} A zoom-in on the stem of the same species, displaying a dense forest of different trichome types. The non-glandular trichomes are generally taller and extend out of the image frame, while the glandular trichomes have a typical height of around $1\,\mathrm{mm}$. \textbf{C.} Close inspection of the short glandular trichomes. The glandular head of the trichomes are each composed of four gland cells surrounding the solvent cavity (boundaries not visible), covered by a cuticle layer.}
    \label{fig:tomato}
\end{figure}

One such defense mechanism that has independently evolved numerous times and which is present in many plants (including most angiosperms) is trichomes \cite{Wang2021AnalysisPlants,Wagner2004NewTrichome}.
Trichomes are small hair-like outgrowths present on the aerial tissues of a plant, consisting of differentiated collections of epidermal cells \cite{Tissier2012GlandularTags, Chang2019SpatiotemporalTomato}. There is an enormous diversity of trichome types according to their shapes, number of constituent cells, and functions \cite{Levin1973TheDefense}. Even for a given species, multiple types of trichomes can be present on the plant; species like tomato can display up to seven different types \cite{Luckwill1943TheTomatoes.}. Trichome sizes range anywhere from a few micrometers to several centimeters \cite{Tissier2012GlandularTags}. However, a commonly applied classification for trichomes distinguishes between whether they possess cells that produce and store specialized metabolites, \emph{i.e.}, \textit{glandular trichomes}, or if they lack these cellular factories, in which case they are called \textit{non-glandular trichomes} (Fig. \ref{fig:tomato}). 
% 
%%%%%%%%%%%%%%%%% What has been studied about trichomes - biochemistry %%%%%%%%%%%%%%%%%%
Trichomes serve multiple purposes in plants, including water retention, temperature regulation, and reflecting light to reduce leaf absorptance \cite{Wagner2004NewTrichome}. However, defense against predation by insects is a primary role of trichomes, where they serve as mechanical and chemical barriers \cite{Kennedy2003TomatoLycopersicon}. Strong correlations exist between trichome density and reduced insect reproduction, development rates, and oviposition, as well as increased mortality rates \cite{Levin1973TheDefense}.

Current research on trichomes frequently focuses on the biochemical aspects of their defensive capacity, including pathways for specialized metabolite biosynthesis and mechanisms regulating their density and development \cite{Bleeker2012ImprovedRelative, Wang2021AnalysisPlants, Bergau2015TheHabrochaites, Han2022MolecularDevelopment}. These investigations attempt to exploit trichome defense compounds to enhance a crop's resistance to pests through genetic approaches \cite{Glas2012PlantHerbivores}. 
% 
%%%%%%%%%%%%%%%%% What has been studied about trichomes - physical properties %%%%%%%%%%%%%%%%%%
Less attention has been given to the role of trichomes' physical properties in plant defense \cite{Bar2019PlantModel}. Direct mechanical stimuli to trichomes are known to induce chemical signals originating at the trichome's basal cells in both \textit{Arabidopsis thaliana} trichomes and long digital (type I and II) trichomes of the cultivated tomato \textit{Solanum lycopersicum}. This is evidence that trichomes in these species serve as sophisticated mechanosensitive switches that activate immune responses against predation \cite{Zhou2017TheSwitch, Matsumura2022MechanosensoryThaliana, Sun2024DifferentTomato}. Specialized hook trichomes can immobilize insects \cite{Bustamante2017ForcesBugs}, while the rupture by insects of type VI trichomes in the \textit{Solanacae} family releases a toxic glandular solvent that impedes insect motion \cite{Snyder1984LeafMites, Stringer1946AAphids, GIBSON1971GlandularSpecies}. However, the physical understanding of how glandular trichomes rupture under stress due to insect locomotion still remains elusive.

\pagebreak
%%%%%%%%%%%% Broad summary paragraph of the work, to transition into results %%%%%%%%%%%%%
In this work, we study the mechanical rupture of type VI glandular trichomes in two species of tomato, modern, cultivated tomato plants \textit{Solanum lycopersicum} (cultivar) and the ``hairy tomato", a wild type relative from the same genus, \textit{Solanum habrochaites}, accession PI127826 (wild). The wild accession has a higher trichome density, different relative abundances of trichome types, and produces more glandular exudate in their trichomes than cultivar tomatoes \cite{Snyder1985TrichomesHybrids}. Type VI trichomes are biosynthesis factories consisting of four glandular cells atop one intermediate cell and a single stalk cell \cite{Tissier2017PlantCavities}. We study type VI trichomes in particular because they are abundant on tomato stems and leaves, and they produce and store the vast majority of glandular solvent in this plant \cite{Bergau2015TheHabrochaites}. To simulate the effect of the impact of an insect's leg in a reproducible fashion, we fabricated glass microcapillary tubes into precise force sensors (see Materials and Methods). We measured, for the first time, the force to rupture glandular trichomes and how rapidly the glandular solvent is released upon rupture. We then observed \textit{in situ} insect-triggered glandular trichome rupture in experiments with the L2 larvae of Western flower thrips (\textit{Frankliniella occidentalis}), demonstrating the plant's effective defense strategy.

%%%%%%%%%%%%%%%%%%%%%%%%%%%%%%% Results %%%%%%%%%%%%%%%%%%%%%%%%%%%%%%%%%%%%%%%%%
\section*{Mechanically Weak Cell Junction Leads to Rapid Rupture of Glandular Trichomes}

\begin{figure}[!tbph]
    \centering
    \includegraphics[width=1\textwidth]{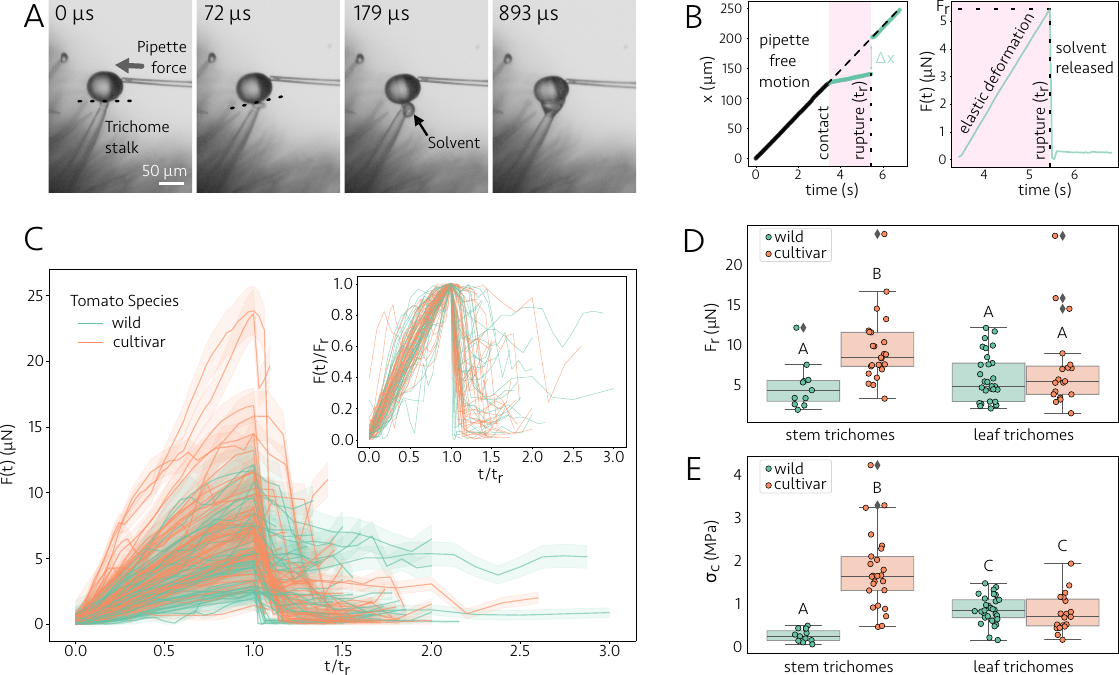}
    \caption{\textbf{Mechanics of glandular trichome rupture.} \textbf{A.} Wild tomato trichome rupture occurs very rapidly upon the application of small amounts of force by a micropipette force sensor, with labeled times beginning from the last frame before rupture. Dashed line denotes the mechanically weak junction between the glandular and intermediate cells where rupture originates. These frames are from a high-speed video filmed at 28,000 fps (Supplementary Video 1). \textbf{B.} A representative experimental curve of the pipette's vertical center of mass location over time (left), and the corresponding force loading curve for trichome rupture (right). Two regimes are observed: pipette free motion before contact with the trichome head (black), and motion after the pipette makes contact (green). Note that after rupture the curve returns to the same slope as the unobstructed curve. \textbf{C.} Force loading curves for all $N=84$ measured ruptures of WT and cultivar tomato trichomes. Confidence intervals bounded by one standard deviation in uncertainty are shaded around the measured curves. The time axis for each curve is normalized by its measured time to rupture, so that each curve has its peak force (the force to rupture $F_r$) at rescaled time $t/t_r = 1$. Inset: the loading curves collapse when both axes are normalized by their values at the point of rupture, demonstrating a universal fracture mechanism. Confidence intervals were removed for visual clarity. \textbf{D.} The rupture forces for cultivar stem trichomes are significantly larger than for other species/trichome location pairs (largest p-value between pairs in Tukey's Honestly Significant Difference (HSD) test was $p=0.001$). \textbf{E.} The critical stress to rupture trichomes is also larger for cultivar stem trichomes than other species/location pairs with even higher significance, and the wild stem trichomes rupture at a significantly smaller critical stress than leaf trichomes. Above each box plot in \textbf{D} and \textbf{E}, different letters indicate statistically significant differences between groups according to Tukey's HSD test ($p < 0.02$), while groups sharing the same letter are not significantly different.}
    \label{fig:rupture-force}
\end{figure}

%%%%%%%%%%%%%%% Figure 2 Description &&&&&&&&&&&&&&&&&&&&&&&&&&&&&&
Rupture of type VI glandular trichomes in wild and cultivar tomatoes was performed using micropipette force sensors under a bright-field microscope (see Materials and Methods). In practice, this consisted of loading either a stem or leaflet sample under the microscope, bringing the trichomes at the edge of the sample in focus, and then moving the micropipette tip into a frame using a micromanipulation stage. The tip would then be translated at a constant velocity towards the trichome glandular head until they made contact, where there would be a short period of force loading before the head ruptured. This rupture occurred extremely rapidly, releasing the contained solvent faster than the highest frame rate we could capture under the microscope with a ``standard" scientific camera (up to 1000 fps). However, with the aid of high-speed imaging (up to 28,000 fps; see Fig. \ref{fig:rupture-force}A. Also see Materials and Methods), we discovered that the solvent is released from its interior cavity in under $1\, \mathrm{ms}$ after the onset of detectable rupture. This solvent reliably exits in the form of a droplet that wets the trichome stalk and grows in size until all of the solvent has been released. There are no detectable jetting or spray phenomena, which indicates that surface tension dominates the fluid dynamics of the problems.
This can be quantified by estimating the non-dimensional numbers:
\be
Re= \rho \, U \, L / \mu , \quad We = \rho \, U^2 \, L / \sigma, \quad Bo= \rho \, g \, L^2 / \sigma,
\ee

where, $\rho$ and $\sigma$ are the density and surface tension of the liquid, respectively, estimated to be close to water, $g$ is the gravitational acceleration and $U$ and $L$ are characteristic velocity and length estimated from the observations. The Reynolds number, $Re$, compares the inertial to viscous forces. The Weber number, $We$, measures the inertial to surface tension forces, and the Bond number, $Bo$, shows the ratio of gravitational to surface tension forces. For the range of parameters in our work, we estimated these values to be $Re\sim \mathcal{O}(0.01-0.1)$, $We \sim \mathcal{O}(10^{-3}-10^{-2})$, and $Bo \sim \mathcal{O}(10^{-5}-10^{-4})$. This places the secretion of glandular fluids at a low $Re-We-Bo$ regime, which corresponds to a negligible effect of gravity and inertia and insufficient energy to produce jetting~\cite{Lohse2021FundamentalPrinting,Challita2024FluidNature}.

%%%%%%%%%%%%%%%%%%%%%%%%%%%% Figure 2 %%%%%%%%%%%%%%%%%%%%%%%%%%%%%
The localization of rupture was highly repeatable ($N=84$), originating at the boundary connecting the glandular trichome head and the intermediate cell of the trichome stalk. Although the two tomato species we studied have different glandular head morphologies, the rupture location and the speed at which the rupture occurs are unaffected (Supplementary Videos 1 and 2). This is in agreement with previous research that proposed this junction as a ``micro-abscission zone" or breaking point for the release of the interior solvent, with rupture occurring under the application of ``a light touch" \cite{Bergau2015TheHabrochaites}. Here, we provide direct experimental observations of this phenomenon, probing the full dynamics of the solvent release. Additionally, until this point, the exact force needed to rupture the glandular head was unknown, aside from the assumption that small pest insects were capable of causing this rupture.
By rupturing a trichome head with a well-calibrated micropipette force sensor (spring constant $k$), we could extract the applied force at all times from the pipette's deflection $\Delta x$. As the pipette translates at a constant speed and pushes against the trichome head, it bends upon contact. We calculate $\Delta x$ from the center of mass location of the micropipette tip $x$ as the difference between where it would be if it had continued moving unobstructed and where we actually measure it to be during contact with the glandular head. The applied force is then 
\be \label{eq:force}
F(t)=k\,\Delta x(t),
\ee
where $t$ is time. A representative experimental curve for one trichome rupture of the pipette tip's vertical center-of-mass location $x$ and the corresponding force loading curve $F(t)$ from Eq. \ref{eq:force} are shown in Fig. \ref{fig:rupture-force}B. The two constant-slope regimes of the pipette $x$-position during its free motion and after trichome contacts are indicative of its constant translation speed during these two intervals, with a shallower slope and thus slower speed after contact during force loading. After glandular rupture, the pipette returns to moving at the same speed as prior to the contact.

The collection of all force loading curves from the ruptures of both wild and cultivar tomato trichomes, sourced from the stems and leaves of the plants, is shown in Fig. \ref{fig:rupture-force}C ($N=84$). Despite rupturing after various times and a range of forces ($1-24\,\mathrm{\mu N}$), the rupture processes were all quite similar. Specifically, the force curves consist of a period of elastic loading, indicated by a positive linear slope in Fig. \ref{fig:rupture-force}C, followed by sudden failure, indicated by the steep drop in force to nearly zero due to trichome rupture removing contact between the glandular head and the pipette. Curves that do not drop entirely to zero force are the result of the glandular fluid sticking to the pipette head, providing a small resistive force to its motion. The consistency in rupture dynamics is demonstrated by rescaling the forces for each measurement by their peak force $F_r$, shown in the inset of Fig. \ref{fig:rupture-force}C. The observed scaling collapse indicates a universal fracture mechanism underlying the rupture, namely brittle fracture. Brittle fractures are characterized by a lack of substantial plastic deformation prior to failure and propagate at extremely high speeds in a material, both of which are properties we observed in our rupture experiments \cite{Field1971BrittleApplication,Barsom1999FractureMechanics, Divoux2024Ductile-to-brittleQuestions}.

The maxima of each of the measured force loading curves define the rupture force $F_r$. We compare the rupture forces between the two tomato species and the trichome's location on the plant in the box plots of Fig. \ref{fig:rupture-force}D. Aggregating all of the rupture force data for each species and location on the plant, we find a mean rupture force of $F_r=7\pm 4\,\mathrm{{\mu N}}$. However, the cultivar tomato stem trichomes had a significantly higher rupture force than other species/location pairs (p=0.001 or smaller for all pairs following a two-way ANOVA test and post-hoc Tukey HSD test). One hypothesis for the difference between cultivated and wild tomatoes in our force to burst type VI trichomes result is that the chemical makeup of the junction point at the neck cells is different between these two species. The difference in cuticle composition between the species, in addition to their known differences in glandular cell morphologies, might also play a role. This motivates future biochemical studies.

Previous research on type VI trichomes in tomatoes has observed a distinct boundary at the junction between the thick glandular cell wall and thinner intermediate cell wall \cite{Bergau2015TheHabrochaites}. We posit that external forces concentrate stress at this two-dimensional junction, triggering trichome rupture. To compute the stress concentration due to trichome bending, we model the trichome as an Euler-Bernoulli cantilever beam \cite{Ochsner2021EulerBernoulliTheory}. Treating the trichome junction as a circular cross-section with measured radius $R$, its axial second moment of area is
\be
I_z = \frac{\pi R^4}{4}.
\ee
As the glandular head rotates at the moment of rupture (Fig. \ref{fig:rupture-force}A), it is useful to quantify the torque %(or ``bending moment" in the engineering literature) 
experienced at the junction. Under the vectorial force $\boldsymbol{F}$ applied a distance vector $\boldsymbol{r}$ from the central axis of the junction, the trichome experiences a torque
\be\label{eq:torque}
\bm{\tau} = \boldsymbol{r} \times \boldsymbol{F}=\left|\boldsymbol{r}\right|\left|\boldsymbol{F}\right|\sin{\theta}, 
\ee
where $\theta$ is the angle between $\boldsymbol{F}$ and $\boldsymbol{r}$. Measurements of the torque to rupture (Fig. \ref{fig:rupture_torque} in Appendix A) showed similar results to that of $F_r$, with a significantly higher torque to rupture cultivar stem trichomes than other species/location pairs. This indicates that our observed higher rupture force for cultivar stem trichomes is not explained by experimental variability in the pipette's angle of force application, or by variations in the glandular head size among species/location pairs.

Under the applied torque $\boldsymbol{\tau}$ from Eq. \ref{eq:torque}, classical beam theory maintains that the junction's stress field varies linearly with distance from its center, and is maximal at the boundary of the interface \cite{Ochsner2021EulerBernoulliTheory}. Denoting the torque to rupture by $\tau_{r}$, the critical stress on the boundary (distance $R$ from the center) at the moment of trichome rupture is then given by
\be \label{eq:stress}
\sigma_{c}=\frac{R\tau_{r}}{I_z}.
\ee
The critical stress was calculated using Eq. \ref{eq:stress} for each of the experimental trichome ruptures, see Fig. \ref{fig:rupture-force}E. The cultivar stem trichomes rupture at a critical stress that is significantly higher than other pairs. A new significant difference appeared for wild stem trichomes, which rupture at a smaller critical stress than the other pairs. These stress calculations, which account for the torque applied by the pipette regardless of angle, demonstrate that the rupture mechanism and differences between species are governed by intrinsic structural properties rather than just the magnitude of applied force. The significant differences in critical stress between stem and leaf trichomes may also align with known biochemical variations in the solvent content in type VI trichomes across plant tissues \cite{Xu2019SlMYC1Cells}. An alternative hypothesis is that differences in the solvent/gland head volume ratio may provide different pressures on the junction, affecting the rupture energy at the junction.

%%%%%%%%%%%%%%% Figure 3 Description &&&&&&&&&&&&&&&&&&&&&&&&&&&&&&

\begin{figure}[h!]
    \centering
    \includegraphics[width=1\textwidth]{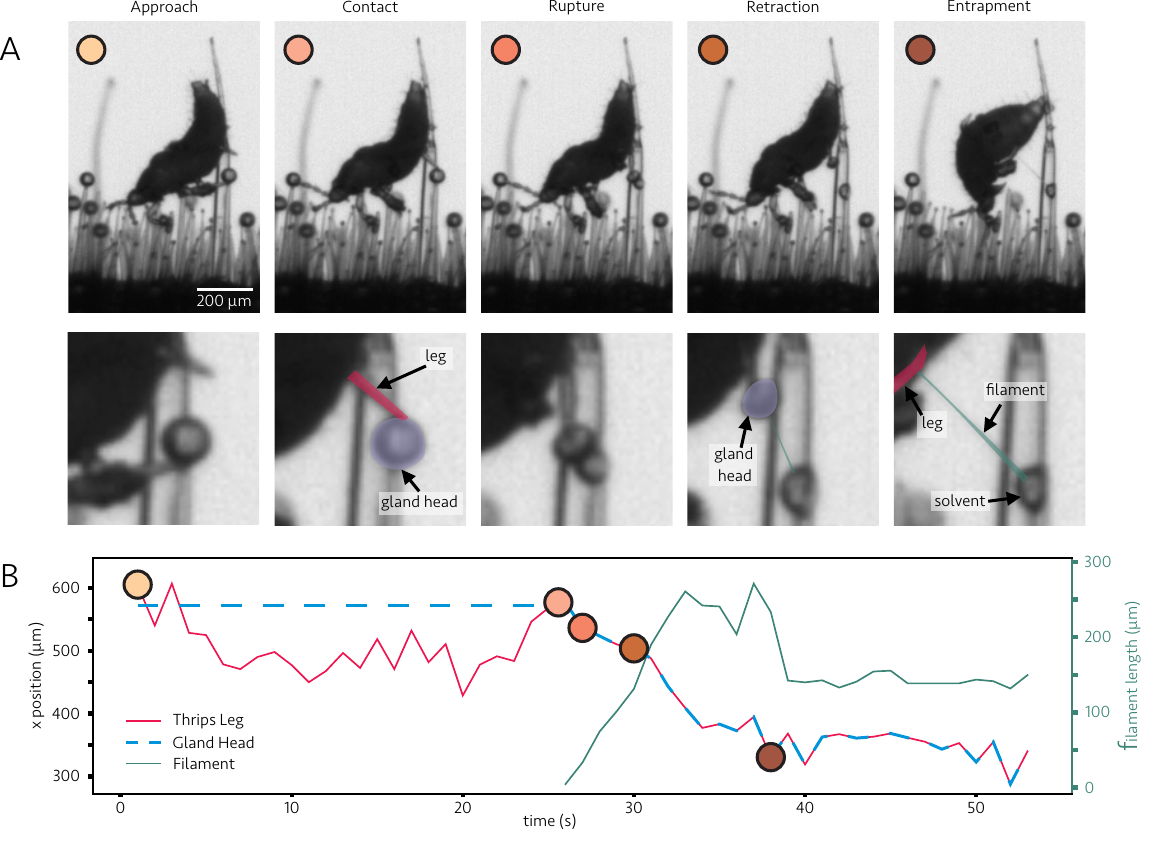}
    \caption{\textbf{Thrips nymphs rupture wild tomato trichomes.} \textbf{A.} Frames from a microscopy video before, during, and after a thrips leg causes glandular trichome head rupture. Zoomed-in panels are shown below each frame, with illustrative labels for the leg, gland head, and solvent filament formed afterward. The colors assigned to the frames indicate the time of the experiment in the leg position vs. time plot of \textbf{B.} The plot also tracks the formation and length of the filament length over the course of the video, as the insect struggles to move. See Supplementary Video 3.}
    \label{fig:insect}
\end{figure}

\section*{Glandular Trichome Fluid Acts as a Mechanical Barrier to Insects}
To validate the biological relevance of our laboratory measurements of the mechanics of tomato trichome rupture, we observe the \textit{in situ} physical interactions of a pest with these trichomes. Insects are known to be impacted in various ways by the release of plant exudates, such as from glandular trichomes \cite{Glas2012PlantHerbivores}. In particular, adult Western flower thrips (\textit{Frankliniella occidentalis}), a common pest of tomato plants, are known to have an avoidance response to the presence of volatile organic compounds produced by these glandular trichomes \cite{Delphia2007InductionThrips}. The larval stages of thrips lack wings and develop entirely on the leaves of tomato plants \cite{Steenbergen2018ThripsCrops}. Hence, we chose to investigate trichome dynamics with the L2-stage larva of the Western flower thrips (see Materials and Methods for details of their rearing). 

A black-and-white bright-field microscopy video of a thrips on a section of wild tomato stem containing many type VI trichomes is presented as Supplementary Video 3 in the Supplementary Data. Fig. \ref{fig:insect}A displays several frames, before, during, and after a thrips leg causes glandular trichome head rupture. We observed the glandular head become stuck to the leg, which then became held by a sugary filament secreted from the trichome after rupture. Over time, the thrips accumulate increasingly more solvent on its body, impeding its motion. This filament appears viscoelastic based on our observation of rupture experiments with micropipettes. The chemical makeup of the solvent varies between species, but a major component in both that contributes to their solvent's stickiness is volatile terpenes \cite{Kortbeek2023GeneticTrichomes, Xu2019SlMYC1Cells}. These are stored in a lipid environment between the glandular cells, and are well-known to have insecticidal effects and cause physical entrapment of the insects \cite{Puterka2003Structure-FunctionEsters, Simmons2004EntrapmentSpecies, Levin1973TheDefense}. The fluid properties of these trichome solvents may function similarly to other plant trapping fluids, like the viscous digestive liquid in pitcher plant traps, where specific physical properties are essential for effectively capturing and retaining prey \cite{Gaume2007APlants}. We have estimated the viscosity of the trichome solvent using measurements of capillary flows inside the glass micropipette (see Appendix B). Our preliminary results show the viscosity of the liquid appears much larger than that of water ($\mu \sim \mathcal{O}(0.1-1)$ Pa.s), similar to that measured for the pitcher plants' digestive liquids. A detailed rheological characterization of these trichome secretions, using shear and extensional rheology, would help reveal the physical mechanisms underlying their trapping success. 

%%%%%%%%%%%%%%%%%%%%%%%%%%%%%%% Discussion %%%%%%%%%%%%%%%%%%%%%%%%%%%%%%%%%%%%%%%%%%%%%%

%%%%%%%%%%%%%%%%%%%%%%%% Insect Sizes Table %%%%%%%%%%%%%%%%%%%%%%%%%%%%%%
\section*{Conclusion, Discussion \& Outlook}
The consistency of the trichome rupture force loading curves in Fig. \ref{fig:rupture-force}C is evidence of a universal fracture mechanical process underlying the dynamics in these species. In this work, we modeled the type VI tomato trichome as a cantilever beam with a weak plane at a specific location, namely the junction between the glandular cells and the intermediate cell. The material properties of the cells on either side of this junction are not necessarily the same. This is similar to classical beam fracture in the mechanics of slender structures, and finite element method simulations could elucidate more about the fundamental material properties at the junction that allow for such robust sequestering of glandular solvent \cite{Anderson2017FRACTUREEdition}. 

Morphologically, the concentration of stress at this junction, which leads to such consistent rupture dynamics, may be a result of the developmental process of the storage cavity itself. Young glandular cells with cell walls that are initially in contact undergo localized lysis, gradually separating from the middle lamella and changing their chemical properties to make space for the storage cavity \cite{Tissier2017PlantCavities}. Pectin demethylation of the cell wall is known to be a critical part of cell wall separation in tomatoes \cite{Bergau2015TheHabrochaites}, and as the cavity expands to connect with the intermediate cell, it is possible that an overabundance of demethylated pectin becomes localized along the junction, providing a weak plane for mechanical rupture. Additional study of the microscopic structures of trichome glandular heads could shed light on these predictions. %Such simulations may benefit from our measurements of the the bending modulus for nonglandular and glandular trichome types II, I/IV, and VI, see Fig. \ref{fig:bending_modulus} in the Supplementary Data.

Our mechanical analysis confirms that the physical properties of trichome rupture are complementary to their already well-established chemical roles in plant defense against insects \cite{Tissier2012GlandularTags,Steenbergen2018ThripsCrops, Kortbeek2018EndogenousInsects}. The observation that L2 stage thrips larvae can rupture trichomes provides a useful size threshold for understanding this defense mechanism. Previous studies have shown similar rupturing by larger insects like cotton bollworms (\textit{Helicoverpa armigera}) and aphids in \textit{Solanum polyadenium} \cite{Simmons2004EntrapmentSpecies,GIBSON1971GlandularSpecies}. 

\begin{table}[h!]
\centering
\begin{tabular}{|l|l|l|l|}
\hline
\textbf{Pest Species} & \textbf{Life Stage} & \textbf{Size (mm)} & \textbf{Reference} \\
\hline
\rowcolor{gray!15} Tobacco Hornworm (\textit{Manduca sexta}) & Final instar larva & 81.3 & \cite{Madden1945BiologyDistrict} \\
\rowcolor{gray!15} Cutworm (\textit{Agrotis ipsilon}) & Final instar larva & 50.0 & \cite{Capinera2001HandbookPests} \\
\rowcolor{gray!15} Brown Stink Bug (\textit{Euschistus servus}) & Adult & 11.0--15.0 & \cite{Capinera2001HandbookPests} \\
\rowcolor{gray!15} Tomato Leafminer (\textit{Tuta absoluta}) & Adult & 5.3--6.2 & \cite{Bajracharya2018LifeNepal} \\
\rowcolor{gray!15} Potato Aphid (\textit{Macrosiphum euphorbiae}) & Adult & 2.1--4.0 & \cite{Capinera2001HandbookPests} \\
\rowcolor{gray!15} Green Peach Aphid (\textit{Myzus persicae}) & Adult & 1.8--2.1 & \cite{Capinera2001HandbookPests} \\
\rowcolor{gray!15} Western Flower Thrips (\textit{Frankliniella occidentalis}) & Adult & 1.2--1.9 & \cite{Capinera2001HandbookPests} \\
\rowcolor{gray!15} Greenhouse Whitefly (\textit{Trialeurodes vaporariorum}) & Adult & 1.0--2.0 & \cite{Capinera2001HandbookPests} \\
\rowcolor{gray!15} Silverleaf Whitefly (\textit{Bemisia tabaci}) & Adult & 1.0-1.3 & \cite{Capinera2001HandbookPests} \\
\rowcolor{gray!15} Western Flower Thrips$^*$ (\textit{Frankliniella occidentalis}) & L2 larva & 0.7--0.8 & \cite{Steenbergen2018ThripsCrops} \\
\rowcolor{gray!15} Greenhouse Whitefly (\textit{Trialeurodes vaporariorum}) & Fourth instar larva & 0.75 & \cite{Capinera2001HandbookPests} \\
\hline
Two-spotted Spider Mite (\textit{Tetranychus urticae}) & Adult & 0.4--0.5 & \cite{Capinera2001HandbookPests} \\
Silverleaf Whitefly (\textit{Bemisia tabaci}) & First instar larva & 0.27 & \cite{Capinera2001HandbookPests} \\
Tomato Russet Mite (\textit{Aculops lycopersici}) & Adult & 0.15--0.2 & \cite{Capinera2001HandbookPests} \\
\hline
\end{tabular}
\caption{\textbf{Size comparison of common tomato pests, sorted by maximum body length.} The shaded region indicates pests of size equal to or larger than L2 stage Western Flower Thrips ($^*$), which were observed to rupture type VI glandular trichomes. This suggests that all pests in the shaded region are likely capable of rupturing these defensive structures through normal movement, while the unshaded pests may be unable to rupture them.}
\label{tab:pest_sizes}
\end{table}

As shown in Table \ref{tab:pest_sizes}, many common tomato pests are equal to or larger in linear size than L2 thrips larvae (shaded region), suggesting they could also trigger trichome rupture through normal movement. This indicates that trichomes may have evolved to rupture at forces corresponding to insects above a critical size threshold, effectively acting as a passive mechanical filter against herbivores. Note that the distribution of leg forces in insects may widely vary, depending on the insect anatomy and environmental stresses~\cite{Full1995StaticMeasurements, Delcomyn2004InsectRobotics, Holmes2006TheChallenges} and furthermore locomotion forces do not scale linearly with insect size, particularly for the smallest pests. Hence, linear size serves as a simple indicator for force distribution, and further research is needed to fully characterize the relationship between insect biomechanics and trichome rupture thresholds.

In summary, we have measured the force to rupture the glandular heads of type VI tomato trichomes, in both wild type and cultivar species. The rupture consistently originated at the junction between the glandular cells and the intermediate cell, in agreement with previous research on trichome development \cite{Bergau2015TheHabrochaites}. The rapidity at which all of the solvent contained in the glandular heads is released, occurring in less than one millisecond, adds trichome glandular rupture to a rare list of extremely rapid plant motions like explosive seed and spore dispersal and elastic buckling mechanisms in some carnivorous plants, contrasting with plants' typically slow, imperceptible movements \cite{forterre2013slow, Vincent2011Ultra-fastTraps, Box2024UncoveringCucumber}. The rupture force measurements ranged from $1.4$ to $23.8\, \mathrm{\mu N}$ (mean $7\pm 4\,\mathrm{{\mu N}}, N=84$), with trichomes from the stems of cultivar tomatoes having significantly higher rupture forces than from the leaves or any location on wild plants. This increased mechanical resistance to rupture in cultivar stem trichomes likely compromises their defensive function, since effective pest deterrence relies on readily bursting trichomes that can quickly release their protective compounds upon insect contact. Historically, cultivated tomatoes have been selected mostly on valuable fruit characteristics such as fruit size, fruit consistency, and high yield. Perhaps this process may have inadvertently impacted the physical properties of the stem trichomes in particular. 

Future research should examine additional wild tomato ancestors and characterize the molecular composition of the glandular-intermediate cell junction, to understand the genetic basis of trichome mechanical properties. Such insights into wild tomato defense mechanisms could guide breeding programs to recover robust pest resistance while maintaining the desirable traits of cultivated varieties, potentially reducing reliance on chemical pesticides.
An immediate opportunity to build on this work is a more detailed rheological characterization of the gland-secreted solvent fluids. Our observations indicate that, in addition to the well-known role of volatile chemicals in repelling insects, the solvent also acts as a mechanical barrier by adhering to the animal's legs and forming long filaments. Establishing a link between the glandular contents and the mechanical properties of the solvent (\emph{e.g.}, frequency-dependent viscoelasticity) could offer deeper insights into the evolution of plant defense mechanisms.

%%%%%%%%%%%%%%%%%%%%%%%%%%%%%%% Acknowledgements %%%%%%%%%%%%%%%%%%%%%%%%%%%%%%%%%%%%%%%%%%%%%%
\section*{Acknowledgements}
We are grateful to Roxane Snijders for helping to obtain the thrips for our experiments and for advice on safe practices with them. We thank Yuri Sinzato, Nico Schramma, and Dominic Vella for fruitful discussions. We also thank the staff of the Swammerdam Institute for Life Sciences greenhouses, Harold Lemereis, Ludek Tikovsky, and Kevin van de Brug, for taking care of our tomato plants during the study. We acknowledge Corentin Coulias for providing access to his high-speed camera. M.J. acknowledges the ERC grant no.~"2023-StG-101117025, FluMAB" and NWO (Dutch Research Council) under the MIST program.

%%%%%%%%%%%%%%%%%%%%%%%%%%%%%%% Competing interests %%%%%%%%%%%%%%%%%%%%%%%%%%%%%%%%%%%%%%%%%%%%%%
\section*{Competing interests}
The authors declare no competing interests.

%%%%%%%%%%%%%%%%%%%%%%%%%%%%%%% Author contributions %%%%%%%%%%%%%%%%%%%%%%%%%%%%%%%%%%%%%%%%%%%%%%
\section*{Author contributions}
J.P., L.W., and M.J. designed the research; J.P., L.W. collected data. J.P. analyzed the data and wrote the first draft. A.C. and P.B. provided the biological samples. M.J. acquired funding. All authors discussed the results and contributed to the final manuscript.

%%%%%%%%%%%%%%%%%%%%%%%%%%%%%%% Funding %%%%%%%%%%%%%%%%%%%%%%%%%%%%%%%%%%%%%%%%%%%%%%
% \section*{Funding}

%%%%%%%%%%%%%%%%%%%%%%%%%%%%%%% Data availability %%%%%%%%%%%%%%%%%%%%%%%%%%%%%%%%%%%%%%%%%%%%%%
% \section*{Data availability}

\printbibliography

% \clearpage
% \renewcommand\thefigure{S\arabic{figure}}
% \setcounter{figure}{0}    
% \section*{Supplementary Information}
% \subsection*{Supplementary Movies}
% Supplementary ...\\

\pagebreak
%%%%%%%%%%%%%%%%%%%%%%%% Rupture torque %%%%%%%%%%%%%%%%%%%%%%%%%%%%%%%%%%%
\section*{Materials and methods}
\subsection*{Plant materials and growth conditions}
Tomato plants from the wild accession PI127826 (\textit{S. habrochaites}) and cultivar Moneymaker (\textit{S. lycopersicum}) species were grown in a greenhouse facility that maintained consistent environmental conditions (24 $^\circ$C; 60\% humidity; 16/8h light/dark cycle). Each plant was watered three times per week and fertilized twice per week on a regular schedule. Leaflets and stem tissues were harvested from plants of the same age, at the third leaf down from the shoot apex.

\subsection*{Insect materials}
Western flower thrips (\textit{F. occidentallis}) were raised on cucumber plants in enclosed, temperature-controlled incubators at constant environmental conditions (22 $^\circ$C; 48\% humidity; 16/8h light/dark cycle). For all experiments involving interactions with tomato, individual L2 larvae were collected in $1.5\,\mathrm{mL}$ Eppendorf tubes and transferred into a separate lab space for imaging, which was otherwise devoid of plants to avoid the spread of this invasive pest insect.

\subsection*{Imaging}
Bright-field microscopy was performed with a Nikon TI2 microscope using a halogen light source with a Photometrics BSI Express sCMOS camera. Glandular rupture experiments involving micropipette force sensors utilized a 4x objective lens with a numerical aperture of 0.13. To better visualize the rapidity and localization of rupture, high-speed imaging was performed under the same microscope with a Phantom VEO 640L camera at a sample rate of 28,000 fps and exposure time $33\,\mathrm{{\mu s}}$, using a 10x objective lens with a numerical aperture of 0.3.

\subsection*{Micropipette force sensor manufacture and calibration}
To obtain our flexible micropipette force sensors, we started with a hollow glass capillary tube with $1\,\mathrm{mm}$ outer diameter and $0.75\,\mathrm{mm}$ inner diameter (TW100-4, World Precision Instruments, USA). A P-1000 micropipette puller system (Sutter Instrument, USA) with a pre-installed box filament (FB245B) was used to heat and rapidly pull this tube, which would split into two micropipettes with an approximately $5\,\mathrm{\mu m}$ diameter tip at the end. The machine settings for micropipette pulling were consistent for all micropipettes manufactured (heat 750, pull 70, vel 150, time 200, pressure 100, ramp 709). 

The end of the pulled micropipette was then cut and smoothed using a microforge (MF2, Narishige, Japan). With our experimental setup it was preferable to measure forces parallel to the end of the micropipette, so we created a right-angle bend near the tip of the pipette, also using the microforge. Hence, the straight portion of the pipette before the bend would act as the cantilever, and deflections for force measurements would be measured just before this bend. It was important for deflection analysis of the experimental data to be performed at approximately the same position along the pipette as in the calibration experiment because how much the pipette deflects under load depends on how far along it one measures.

Manufactured micropipettes were calibrated according to the water-droplet and/or two-pipette deflection procedures described in \cite{Backholm2019MicropipetteMicroorganisms}. A custom ImageJ macro for extracting the pipette center of mass position and Python code used to analyze this calibration data according to the procedures in the paper can be found at \cite{pipette_github}. In experiments requiring micropipette translation; they were attached to SensaTex $\mu$Mp-3 micromanipulators that have a high positional resolution. This enabled movement of the micropipette with a $5\, \mathrm{nm}$ resolution, and $100\, \mathrm{nm}$ repeatability. 

Depending on where the pipette was cut, and the location of the bend, the final tip diameter of the pipettes varied somewhat, with a range of 10-$22\,\mathrm{\mu m}$ for the micropipettes used in our experiments. This led to a range of calibrated pipette $k$-values from $k=0.04$ to $k=1.17\, \mathrm{\mu N/\mu m}$, with typical uncertainties of the order less than or equal to 0.01 $\mathrm{\mu N/\mu m}$. For the force measurements using the deflection of these micropipettes, this corresponds to an uncertainty of tens to hundreds of $\mathrm{nN}$, depending on the pipette and the resolution of the imaging camera.

%%%%%%%%%%%%%%%%%%%%%%%%%%%%%%% Supplementary Data %%%%%%%%%%%%%%%%%%%%%%%%%%%%%%%%%%%%%%%%%%%%%%
\pagebreak
\section*{Appendices}

\subsection*{A. Rupture torque measurement }
\begin{figure}[htbp]
    \centering
    \includegraphics[width=0.7\textwidth]{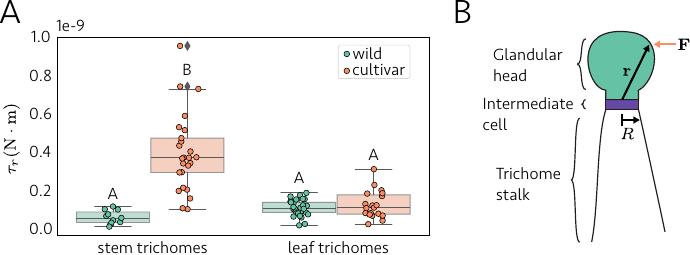}
   \caption{\textbf{A.} Box plots of the torque to rupture tomato type VI glandular trichomes, for the two species and trichome locations on the plant studied in this work. Different letters indicate statistically significant differences between groups ($p < 0.001$), while groups sharing the same letter are not significantly different. \textbf{B.} Schematic of a trichome showing the definitions of $\vec{r}$, $\vec{F}$, and $R$ used in Eqs. \ref{eq:torque} and \ref{eq:stress}.}
    \label{fig:rupture_torque}
\end{figure}

To ensure that our observed higher rupture force for cultivar stem trichomes is not due to the experimental variability in the pipette's angle of force application or the differing morphologies of glandular cells between species, we measured the torque to rupture (see Eq. \ref{eq:torque} and the discussion about it). The results, shown here in Fig. \ref{fig:rupture_torque}, are similar to that of the rupture force in Fig. \ref{fig:rupture-force}D, with a significantly higher torque to rupture cultivar stem trichomes than other stem/location pairs. Despite the two species having different glandular head morphologies, the physical size of the glandular head remains similar, such that observed variations in $\boldsymbol{r}$ arose more from differences in the pipette's location on the head than to the geometry of the head itself.

\subsection*{B. Viscosity estimation}
\begin{figure}[htbp]
    \centering
    \includegraphics[width=1\textwidth]{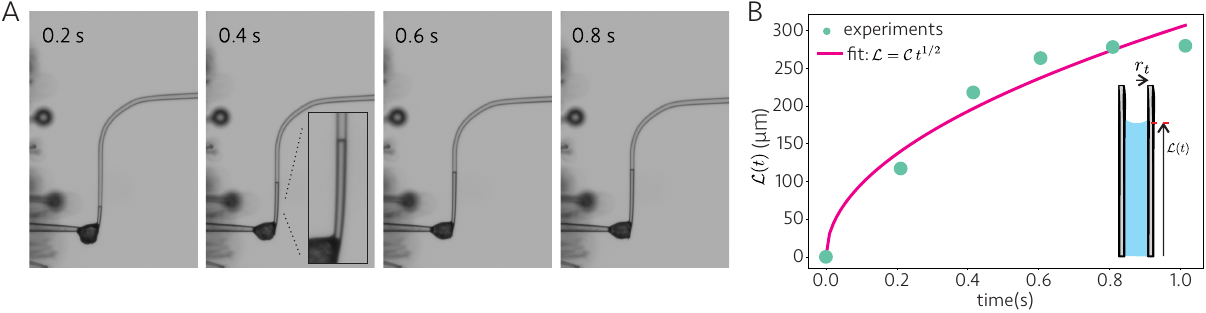}
   \caption{\textbf{A.} Capillary flow and penetration of glandular flow inside the glass micropipette over time. A magnified view is shown in the second panel. \textbf{B.} Variation of penetration length over time, where $\mathcal{C}=304\pm 13$ m/s$^{1/2}$ is the best-fit value of the coefficient in equation~\ref{eq:visc}. Inset shows the schematic highlighting the penetration length $L(t)$ and the radius of the tube, $r_t$.}
    \label{fig:viscosity}
\end{figure}

Using a glass micropipette also enables us to (roughly) measure the solvent's viscosity (Fig. \ref{fig:viscosity}, Supplementary Video 4).
When the glandular head bursts and the fluid is secreted, it rises through the pipette due to capillary action \cite{Washburn1921TheFlow,deGennes2004CapillarityWaves}. This dynamic process allows us to estimate the fluid's viscosity. We apply Washburn's equation, which describes the penetration length of a liquid in a wetting tube:

\be
\mathcal{L}(t)= \mathcal{C} \, t^{1/2};\;\;\;\;\; \mathcal{C}=\sqrt{\frac{\sigma\,r_t\,\mathrm{cos}(\theta)}{2\,\mu}}.
\label{eq:visc}
\ee

Here, $\mathcal{L}$ and $r_t$ represent the penetration length and the radius of the capillary tube, respectively (see the schematic in Figure \ref{fig:viscosity}B). The microcapillary tube gradually increases in radius from $r_1 =4\, \mathrm{\mu m}$ to $r_2 =9\, \mathrm{\mu m}$ over the range of $\mathcal{L}$ values that we plot here. We approximate this tapered cylinder as a cylinder of constant effective radius $r_t$. Since the volume of a tapered cylinder is 
\begin{equation}
\frac{1}{3}\pi h (r_1^2+r_1r_2+r_2^2),
\end{equation}
for a cylinder to preserve this same volume it needs an effective radius $r_t$ of
\begin{equation}
    r_t = \sqrt{\dfrac{r_1^2 +r_1r_2+r_2^2}{3}},
\end{equation}
which for our system gives $r_t = 6.66 \mathrm{{\mu m}}$. Estimating the surface tension $\sigma=72\, \mathrm{{mN/m}}$ and the contact angle $\theta=0^{\circ}$ as for water at room temperature, and our experimental best-fit value for the coefficient $\mathcal{C}=304 \pm 13$ m/s$^{1/2}$ in Eq. \ref{eq:visc}, we estimate the viscosity of the solvent to be $\mu=2.6 \pm 0.1$ Pa·s. Given the lack of experimental repetitions and our uncertainties in the solvent's surface tension, contact angle, etc. we feel confident placing the viscosity in the range $\sim \mathcal{O}(0.1-1)$ Pa·s. 

This value falls within the range measured for certain plant digestive fluids~\cite{Gaume2007APlants} and corresponds to a fluid with significantly higher viscosity than water.
We emphasize that our viscosity estimation using the present method suffers from various simplifications, including unknown surface properties of the liquid, as well as the neglect of other mechanical properties such as elasticity and extensional properties. Moreover, the solvent's highly volatile nature might rapidly influence this biological fluid's rheological properties, as we expect that viscosity rapidly increases as the volatile organics exit the solution and enter the surrounding air. These limitations highlight the need for future studies on the rheology of trichome glandular solvents.

\end{document}